\documentclass[twocolumn,showpacs,preprintnumbers,amsmath,amssymb]{revtex4}
\begin{document}
\title{On the higher order corrections to the Fokker-Planck equation}
\author{A. V. Plyukhin}
 \affiliation{ Department of Physics and Engineering Physics,
 University of Saskatchewan, Saskatoon, SK S7N 5E2, Canada 
}

\date{\today}

\begin{abstract}
The  Rayleigh model of nonlinear Brownian motion is revisited in which the
heavy particle of mass $M$ interacts with ideal gas molecules of mass 
$m\ll M$ via instantaneous collisions.
Using the van Kampen method of expansion of the master equation, 
non-linear corrections to the
Fokker-Planck equation are obtained up to  sixth order in
the small parameter $\lambda=\sqrt{m/M}$, improving earlier results. 
The role and origin of non-Gaussian statistics of the random force
in the corresponding Langevin equation are also discussed.

\end{abstract}


\maketitle

\section{Introduction}
There has been renewed discussion recently of the Rayleigh model
of nonlinear Brownian motion, stimulated primarily by 
findings of new qualitative effects governed by
nonlinear stochastic processes such as 
rectification of fluctuations [1-4], stochastic resonance~\cite{Res}, 
etc. 
In the Rayleigh model 
the heavy Brownian particle of mass $M$
is immersed in an ideal gas of molecules of mass $m\ll M$
and interacts with them through instantaneous elastic collisions.
The gas is assumed to be so rarefied that collisions of its 
molecules with the particle do not affect the distribution of 
incident molecules, and also re-collisions can be neglected.    
In lowest order in the small parameter $\lambda=\sqrt{m/M}$, the velocity
distribution function $f(V,t)$ of the particle satisfies  the
second order Fokker-Planck equation which can be obtained by the 
conventional method, i.e. truncating the Kramers-Moyal expansion 
of the master equation to the first two  terms and evaluating
coefficients integrating the corresponding Langevin 
equation~\cite{Risken,Kampen_book}.
This procedure is consistent if  the random force
in the Langevin equation can be treated as 
white (with negligible correlation time) and Gaussian.
However these two assumptions
may be justified strictly speaking   only in the ultimate limit $\lambda\to 0$.
For finite $\lambda$,  the Kramers-Moyal expansion 
has  generally an infinite number of  non-zero terms, and
more careful analysis is needed  
to study  effects of higher order 
in $\lambda$~\cite{Kampen_book}.
In this case, as was first shown by van Kampen, the differential 
equations for the distribution function do not  have
the form of  the second order Fokker-Planck
equation but rather involve derivatives of order higher than 
two~\cite{Kampen_paper,gen}. 
Although these  higher order equations are linear 
in the distribution function, 
they are sometimes
referred to as nonlinear Fokker-Planck equations because the 
corresponding Langevin equations involve
nonlinear corrections of higher orders in $\lambda$
to the linear damping force. 
These corrections may lead to 
interesting physical consequences, such as 
additional terms in the 
fluctuation spectrum and 
modification of decay rate
coefficients~\cite{Kampen_book}. 
Also, going beyond the lowest order is often  necessary to 
account for  delicate fluctuation-induced phenomena such
as directional  drift in the absence of 
systematic forces.
If the temperatures on the left and right sides of the Rayleigh
particle are different, the particle undergoes  directional movement
even when the pressure on both sides is the same~[2-4]. 
The  systematic average velocity of the particle can be 
calculated as a perturbative solution of 
the Langevin equation with nonlinear corrections to the 
damping force~\cite{I}.  
Alternatively, one can use
the corresponding nonlinear Fokker-Planck equation
or an equivalent set of equations for the moments~\cite{Broeck2}.

To derive a differential equation for the distribution 
function $f(V,t)$ to desirable order
in $\lambda$,  one has to  extract an explicit $\lambda$-dependence 
in the Kramers-Moyal expansion, transforming it 
into the expansion in powers of $\lambda$,
often referred to as the van Kampen expansion (VKE).
The method essentially relies on an assumption of certain
scaling properties of transition rates in the master equation.
For the Rayleigh model an explicit expression for the transition
rates is available and all coefficients in the VKE  
can be found analytically.
Note that another popular exactly solvable model of Brownian motion, 
namely  that of 
a particle coupled to a bath of harmonic 
oscillators, 
is degenerate in the sense that all nonlinear corrections vanish 
identically. 

The Rayleigh model  may be  generalized in many ways 
to study effects of 
asymmetry of the surrounding 
bath~[2-4],
possibility of
non-canonical distributions~\cite{Barkai},
 finite-range interaction~\cite{PS},
strong non-equilibrium fluctuations~\cite{non_equilibrium},  
etc. To analyze these generalizations
it would be of help to compare their predictions  
with the results of the original model. 
In doing  that we have found that the VKE for the Rayleigh
model often appeared in the literature in forms which are either 
not quite accurate or incomplete,  
and lead to nonlinear Fokker-Planck
equations with serious defects. For instance, these equations 
may have a non-Gaussian
stationary solution even when  the Maxwellian distribution is
an exact solution of the original master equation. 
Violation of the equilibrium condition
can lead to  large errors, such as an exponentially
overestimated activation flux in the barrier crossing 
problem~\cite{Hanggi2}.
As was noted by van Kampen in~\cite{Kampen_paper},
the Maxwellian distribution must satisfy 
each term of the VKE for the Rayleigh particle separately.
While the equation of order $\lambda^4$ derived in~\cite{Kampen_paper}
has the Maxwellian stationary solution, the  
equations obtained  later by other authors
 do not always posses this virtue~\cite{Colin,Zhu}. 
The extraction of explicit dependence on $\lambda$ 
for the Rayleigh model is slightly more 
complicated comparing with other applications
of the van Kampen method and requires some care.
One purpose of this paper is to give an accurate 
derivation of
the equations for the velocity distribution of the Rayleigh
particle up to order $\lambda^6$. These equations have 
the Maxwellian stationary solution and preserve positivity of
the distribution function when deviations from equilibrium
are small.  
Another purpose is to discuss some subtle points
relevant to the derivation, such as the origin and consequences of 
non-Gaussian statistics  of the random force in the corresponding
Langevin equation, which 
has not been clearly articulated in the literature so far. 
Although the assumption of Gaussian random force was
criticized in the literature~\cite{Kampen_book},
it is still widely applied to derive
the second order Fokker-Planck equation,
often without proper justification. In the last section
we discuss why the random force is approximately Gaussian in lowest order
in $\lambda$, while corrections of higher orders may be essentially
non-Gaussian.  This fact has important implications which go far beyond
the specific model considered here.

\begin{widetext}

\section{Kramers-Moyal and van Kampen expansions}
The conventional  starting point in the derivation of
the Fokker-Planck equation, as well as equations of
higher order, 
is the master equation for the velocity
distribution function $f(V,t)$
\begin{eqnarray}
\frac{\partial f(V,t)}{\partial t}=\int dV'\Bigl\{
f(V',t)W(V'\to V)-f(V,t)W(V\to V')\Bigr\}.
\end{eqnarray}
It is convenient to express the transition rates 
$W(V_1\to V_2)$ as
a function of the initial state $V_1$ and the transition length
$\Delta V=V_2-V_1$, that is $W(V_1\to V_2)=W(V_1|\Delta V)$. 
Then the master
equation takes a more suggestive form
\begin{eqnarray}
\frac{\partial f(V,t)}{\partial t}
&=&\int d(\Delta V)\Bigl\{
f(V-\Delta V,t)W(V-\Delta V|\Delta V)-f(V,t)W(V|\Delta V)\Bigr\}\nonumber\\
&=&\int d(\Delta V)\Bigl\{\Psi(V-\Delta V,\Delta V)-
\Psi(V,\Delta V)\Bigr\},
\label{aux001}
\end{eqnarray}
where $\Psi(V,\Delta V)\equiv f(V,t)W(V|\Delta V)$.
Making in (\ref{aux001}) the expansion 
\begin{eqnarray}
\Psi(V-\Delta V,\Delta V)=\Psi(V,\Delta V)+
\sum_{n=1}\frac{1}{n!}\left(-\Delta V\frac{\partial}{\partial V}\right)^n
\Psi(V,\Delta V)
\end{eqnarray}
leads immediately to the Kramers-Moyal expansion,
\begin{eqnarray}
\frac{\partial f(V,t)}{\partial t}=
\sum_{n=1}\frac{1}{n!}\left(-\frac{\partial}{\partial V}\right)^n
\Bigl\{a_n(V)f(V,t)\Bigr\}
\label{KME}
\end{eqnarray}
with coefficients $a_n$ given by
\begin{eqnarray}
a_n(V)=\int d(\Delta V)(\Delta V)^n W(V|\Delta V).
\label{a1}
\end{eqnarray} 

If the random force exerted on the particle by
the bath molecules is white and Gaussian, one can show that
only two first terms survive in the Kramers-Moyal expansion, which 
therefore turns into the second order 
Fokker-Planck equation. This can be proved using another
equivalent representation for $a_n$,
\begin{eqnarray}
a_n(V)=\lim_{\tau\to 0}\frac{1}{\tau}
\Bigl\langle(V(t+\tau)-V(t))^n\Bigr\rangle,
\label{a2}
\end{eqnarray}
which can be obtained from (\ref{a1}) writing the  transition rates $W$
in terms of the transition probability $P(V,t|V',t+\tau)$ as
$W(V\to V')=\lim_{\tau\to 0}\frac{1}{\tau}P(V,t|V',t+\tau)$.
Then (\ref{a1}) takes the form 
\begin{eqnarray}
a_n(V)=\lim_{\tau\to 0}\frac{1}{\tau}
\int dV'(V'-V)^nP(V,t|V',t+\tau),\nonumber
\end{eqnarray}
which is  equivalent to (\ref{a2}).

The expression (\ref{a2}) for $a_n$ is often more useful than (\ref{a1}) 
since it does not involves transition rates $W$, which are 
usually unknown. On the other hand the average
$\Bigl\langle(V(t+\tau)-V(t))^n\Bigr\rangle$, appearing in (\ref{a2}),
can be readily found integrating the Langevin equation
\begin{eqnarray}
\dot{V}=A(V)+F(t),
\label{LE}
\end{eqnarray}
where $A(V)$ is the damping force  and $F(t)$ is the 
random force with negligible correlation time,
\begin{eqnarray}
\langle F(t)F(0)\rangle=\Gamma\delta (t).
\end{eqnarray} 
Integrating (\ref{LE}) one gets
\begin{eqnarray}
V(t+\tau)-V(t)=\int_t^{t+\tau}dt'\{A(V(t'))+F(t')\},
\label{aux3}
\end{eqnarray} 
which for small $\tau$ can be consistently approximated as
\begin{eqnarray}
V(t+\tau)-V(t)\approx A(V(t))\tau+\int_t^{t+\tau}dt'F(t').
\label{aux4}
\end{eqnarray}    
This approximation corresponds to
a coarse-grained description with a time resolution $\tau_0$ 
much shorter than
the characteristic time for the relaxation of the particle's 
velocity $\tau_V$ 
and much 
longer than the correlation time for the random force $\tau_F$,
 $\,\,\tau_F\ll\tau_0\ll\tau_V$.
The limit $\tau\to 0$ in (\ref{a2})
means actually $\tau\to\tau_0$.
Therefore, in (\ref{aux3}) one can
neglect time dependence of $A(V(t))$, but not of the random force
which evolves significantly within the integration range 
$\tau\sim\tau_0\gg\tau_F$.

Using the approximation (\ref{aux4}) one can readily find 
$\langle(V(t+\tau)-V(t))^n\rangle$
and then, from Eq. (\ref{a2}), the coefficients
$a_n$.  
For a white  and Gaussian random force
only two first terms survive in the Kramers-Moyal expansion,
\begin{eqnarray}
a_1=A(V),\,\,\,\,\,\,\,\,\, a_2=\Gamma,\,\,\,\,\,\,\,\,\, a_n=0, n>2, 
\end{eqnarray}
which therefore
turns  into the second order 
Fokker-Planck equation
\begin{eqnarray}
\frac{\partial f(V,t)}{\partial t}=
-\frac{\partial}{\partial V}A(V)f(V,t)+
\frac{\Gamma}{2}
\frac{\partial^2}{\partial V^2}f(V,t).
\label{FPE0}
\end{eqnarray}

The result that in the Kramers-Moyal expansion only two first 
terms do not vanish  relies entirely on the assumptions
that the random force in the Langevin equation is 
Gaussian white noise. The above approach does not involve any parameters
controlling applicability of this assumption, and 
the range of validity of the Fokker-Planck equation is difficult 
to analyze.  
This difficulty does not arise in situations when transition rates
in the master equation are known  explicitly. 
In this case one can find $a_n$ directly from Eq.(\ref{a1}) without 
appealing to the Langevin equation and therefore without  
making any assumption about statistical properties of the random force.
Using this method
one finds that the condition
$a_n=0,\,\,\,n>2$ does not  hold invariably, 
and the Kramers-Moyal expansion
contains in general infinite number of terms. 
Nevertheless, it is still possible to get for $f(V,t)$
a  differential equation of finite order  analyzing dependence of
coefficients $a_n$ on the small parameter
$\lambda$ and neglecting terms of higher order in $\lambda$.
The Kramers-Moyal expansion in the form (\ref{KME})
is not appropriate for such a perturbation analysis because 
the dependence on $\lambda$ is implicit in (\ref{KME}). 
Assuming that transition rates have certain scaling properties
with respect to $\lambda$, 
van Kampen modified
the Kramers-Moyal  expansion transforming it 
into the form of the expansion  in powers of 
$\lambda$. Keeping terms up to  order $\lambda^{n}$ in this expansion, 
one obtains  for $f(V,t)$ the differential equation  of order
$n$.

\section{Van Kampen Expansion for the Rayleigh particle}
For the one-dimensional Rayleigh model 
the  transition rate has the form~\cite{Kampen_book} 
\begin{eqnarray}
W(V|\Delta V)=\frac{\nu}{4}\epsilon^{-4}|\Delta V|\,
f_M\left(V+\frac{1}{2}\epsilon^{-2}\Delta V\right),
\label{W}
\end{eqnarray}
where $f_M(v)$ is the Maxwell distribution for the gas molecules,
$\nu$ is the number of particle per unit length, and
\begin{eqnarray}
\epsilon=\sqrt{\frac{m}{M+m}}=\lambda\sqrt{\frac{1}{1+\lambda^2}}.
\end{eqnarray}
The transition rate
has the scaling property
\begin{eqnarray}
\epsilon^2 W(V|\Delta V)=\Phi(V|\epsilon^{-2}\Delta V),
\label{scaling}
\end{eqnarray}
where
\begin{eqnarray}
\Phi(V|\xi)=\frac{\nu}{4}\,|\xi|\,
f_M\left(V+\frac{1}{2}\xi\right).
\label{scaling+}
\end{eqnarray}
To extract an explicit dependence
of $a_n$ on $\lambda$ it is convenient to 
re-write  Eq. (\ref{a1}) as follows
\begin{eqnarray}
a_n(V)&=&\int d(\Delta V)(\Delta V)^n W(V|\Delta V)\nonumber\\
&=&\epsilon^{2n}\int d(\epsilon^{-2}\Delta V)(\epsilon^{-2}\Delta V)^n\,
\Bigl(\epsilon^2 W(V|\Delta V)\Bigr),
\label{aux1}
\end{eqnarray}
or using (\ref{scaling}), 
\begin{eqnarray}
a_n(V)= \epsilon^{2n}\int d\xi \xi^n \Phi(V|\xi).
\end{eqnarray}
Then the Kramers-Moyal expansion takes the form
\begin{eqnarray}
\frac{\partial f(V,t)}{\partial t}=
\sum_{n=1}\frac{1}{n!}\left(-\epsilon^2\frac{\partial}{\partial V}\right)^n
\Bigl\{\alpha_n(V)f(V,t)\Bigr\},
\label{aux2}
\end{eqnarray} 
where 
\begin{eqnarray}
\alpha_n(V)=\int d\xi \xi^n \Phi(V|\xi).
\label{alpha}
\end{eqnarray}
In (\ref{aux2}) the dependence on $\lambda$ enters in two ways.
First, $\epsilon^{2n}$  as a function of $\lambda$ has  the form 
\begin{eqnarray}
\epsilon^{2n}=
\lambda^{2n}\varphi_n(\lambda),
\end{eqnarray}
where
\begin{eqnarray}
\varphi_n(\lambda)=\left(\frac{1}{1+\lambda^2}\right)^n=
1-n\lambda^2+\frac{n(n+1)}{2}\lambda^4
+\cdots
\end{eqnarray} 
Second, the expansion (\ref{aux2}) involves dependence 
on the velocity the particle  $V$  which is obviously a  function 
of $\lambda$. If we restrict ourselves to small-range fluctuations
about the  equilibrium state, it may be  reasonably expected 
from the equipartition theorem that 
the ratio of the particle's typical velocity to that of
surrounding molecules is of order $\sqrt{m/M}$. 
This leads to the second scaling assumption  
\begin{eqnarray}
V=\lambda x
\end{eqnarray}
where the scaled velocity $x\sim \lambda^0$.
Then the next step in the extraction of the 
explicit dependence on $\lambda$
is the expansion of $\alpha_n(V)=\alpha_n(\lambda x)$ near $V=0$,
\begin{eqnarray}
\alpha_n(\lambda x)=\sum_{p=0}\alpha_n^{(p)}\frac{(\lambda x)^p}{p!}
\label{alpha_d}
\end{eqnarray}
where $\alpha_n^{(p)}$ is $p$-th derivative of $\alpha_n(V)$ at
$V=0$. Physically this expansion reflects the fact that
the heavy particle 
is much slower than surrounding light molecules.
The first term with $p=0$  corresponds to
the particle which does not move at all 
(approximation of the infinitely heavy particle, $\lambda\to 0$),
while the next terms successively take into account the finite inertia of 
the particle.

Substitution of the above expansions for $\epsilon^{2n}$ and
$\alpha_n$ in (\ref{aux2}) 
gives finally the desirable van Kampen expansion in powers of $\lambda$    
\begin{eqnarray}
\frac{\partial f(x,t)}{\partial t}=
\sum_{n=1}\frac{(-1)^n}{n!}
\lambda^n\varphi_n(\lambda)
\sum_{p=0}
\frac{\lambda^{p}}{p!}
\alpha_n^{(p)}
\frac{\partial^n}{\partial x^n}
\Bigl(x^p f(x,t)\Bigr).
\label{VKE}
\end{eqnarray}
This form of the VKE is slightly different 
from that one usually finds in the literature
and has the advantage
that the dependence on $\lambda$ is entirely contained in the product 
$\lambda^{n+p}\varphi_n(\lambda)$, whereas
the coefficients $\alpha_n^{(p)}$ are $\lambda$-independent.

The VKE~(\ref{VKE}) can be written in the form
\begin{eqnarray}
\frac{\partial f(x,t)}{\partial t}=\sum_{k=1}^{\infty}
\frac{\partial^k}{\partial x^k}S_k(x,\lambda)f(x,t),
\end{eqnarray}
where the first six coefficients are
\begin{eqnarray}
\label{SSs}
&&S_1(x,\lambda)=
-(\lambda-\lambda^3+\lambda^5+\cdots)\times\\
&&
\Bigl\{ \alpha_1^{(0)}+\lambda\alpha_1^{(1)}x+
\frac{1}{2!}\lambda^2\alpha_1^{(2)}x^2+
\frac{1}{3!}\lambda^3\alpha_1^{(3)}x^3+
\frac{1}{4!}\lambda^4\alpha_1^{(4)}x^4+
\frac{1}{5!}\lambda^5\alpha_1^{(5)}x^5+\cdots
\Bigr\},\vspace{2cm}\nonumber\\
&&S_2(x,\lambda)=\frac{1}{2!}
(\lambda^2-2\lambda^4+3\lambda^6+\cdots)\times\nonumber\\
&&
\Bigl\{\alpha_2^{(0)}+\lambda\alpha_2^{(1)}x+
\frac{1}{2!}\lambda^2\alpha_2^{(2)}x^2+
\frac{1}{3!}\lambda^3\alpha_2^{(3)}x^3+
\frac{1}{4!}\lambda^4\alpha_2^{(4)}x^4+
\cdots
\Bigr\},\nonumber\\
&&S_3(x,\lambda)=
-\frac{1}{3!}
(\lambda^3-3\lambda^5+\cdots)
\Bigl\{\alpha_3^{(0)}+\lambda\alpha_3^{(1)}x+
\frac{1}{2!}\lambda^2\alpha_3^{(2)}x^2+
\frac{1}{3!}\lambda^3\alpha_3^{(3)}x^3+\cdots
\Bigr\},\nonumber\\
&&S_4(x,\lambda)=
\frac{1}{4!}
(\lambda^4-4\lambda^6+\cdots)
\Bigl\{\alpha_4^{(0)}+\lambda\alpha_4^{(1)}x+
\frac{1}{2!}\lambda^2\alpha_4^{(2)}x^2+\cdots
\Bigr\},\nonumber\\
&&S_5(x,\lambda)=
-\frac{1}{5!}
(\lambda^5+\cdots)
\Bigl\{\alpha_5^{(0)}+\lambda\alpha_5^{(1)}x+\cdots
\Bigr\},\nonumber\\
&&S_6(x,\lambda)=
\frac{1}{6!}
(\lambda^6+\cdots)
\Bigl\{\alpha_6^{(0)}+\cdots
\Bigr\}\nonumber.
\end{eqnarray}
It is usually more convenient to write the result in the form
of the expansion in powers of $\lambda$, 
collecting in (\ref{SSs}) terms of the same order,
\begin{eqnarray}
\frac{\partial f(x,t)}{\partial t}=\sum_{k=1}^{\infty}\lambda^k
D_kf(x,t).
\label{myexpansion}
\end{eqnarray}
The first six differential operators $D_k$ are
\begin{eqnarray}
\label{Ds}
&&D_1=-\alpha_1^{(0)}\frac{\partial}{\partial x},\\
&&D_2=-\alpha_1^{(1)}\frac{\partial}{\partial x}x+
\frac{1}{2}\alpha_2^{(0)}\frac{\partial^2}{\partial^2 x},\nonumber\\
&&D_3=
\alpha_1^{(0)}\frac{\partial}{\partial x}
-\frac{1}{2}\alpha_1^{(2)}\frac{\partial}{\partial x}x^2+
\frac{1}{2}\alpha_2^{(1)}\frac{\partial^2}{\partial^2 x}x
-\frac{1}{6}\alpha_3^{(0)}\frac{\partial^3}{\partial^3 x},\nonumber\\
&&D_4=
\alpha_1^{(1)}\frac{\partial}{\partial x}x
-\frac{1}{6}\alpha_1^{(3)}\frac{\partial}{\partial x}x^3
-\alpha_2^{(0)}\frac{\partial^2}{\partial x^2}
+\frac{1}{4}\alpha_2^{(2)}\frac{\partial^2}{\partial x^2}x^2
-\frac{1}{6}\alpha_3^{(1)}\frac{\partial^3}{\partial x^3}x
+\frac{1}{24}\alpha_4^{(0)}\frac{\partial^4}{\partial x^4},\nonumber\\
&&D_5=
-\alpha_1^{(0)}\frac{\partial}{\partial x}
+\frac{1}{2}\alpha_1^{(2)}\frac{\partial}{\partial x}x^2
-\frac{1}{24}\alpha_1^{(4)}\frac{\partial}{\partial x}x^4
-\alpha_2^{(1)}\frac{\partial^2}{\partial x^2}x
+\frac{1}{12}\alpha_2^{(3)}\frac{\partial^2}{\partial x^2}x^3
+\frac{1}{2}\alpha_3^{(0)}\frac{\partial^3}{\partial x^3}\nonumber\\
&&
-\frac{1}{12}\alpha_3^{(2)}\frac{\partial^3}{\partial x^3}x^2
+\frac{1}{24}\alpha_4^{(1)}\frac{\partial^4}{\partial x^4}x
-\frac{1}{120}\alpha_5^{(0)}\frac{\partial^5}{\partial x^5},\nonumber\\
&&D_6=
-\alpha_1^{(1)}\frac{\partial}{\partial x}x
+\frac{1}{6}\alpha_1^{(3)}\frac{\partial}{\partial x}x^3
-\frac{1}{120}\alpha_1^{(5)}\frac{\partial}{\partial x}x^5
+\frac{3}{2}\alpha_2^{(0)}\frac{\partial^2}{\partial x^2}
-\frac{1}{2}\alpha_2^{(2)}\frac{\partial^2}{\partial x^2}x^2
+\frac{1}{48}\alpha_2^{(4)}\frac{\partial^2}{\partial x^2}x^4\nonumber\\
&&
+\frac{1}{2}\alpha_3^{(1)}\frac{\partial^3}{\partial x^3}x
-\frac{1}{36}\alpha_3^{(3)}\frac{\partial^3}{\partial x^3}x^3
-\frac{1}{6}\alpha_4^{(0)}\frac{\partial^4}{\partial x^4}
+\frac{1}{48}\alpha_4^{(2)}\frac{\partial^4}{\partial x^4}x^2
-\frac{1}{120}\alpha_5^{(1)}\frac{\partial^5}{\partial x^5}x
+\frac{1}{720}\alpha_6^{(0)}\frac{\partial^6}{\partial x^6}.\nonumber
\end{eqnarray}
These formulas are valid not only for the original 
Rayleigh model, but also for  
asymmetric models~[2-4] 
when properties of the bath 
on the left and on the right of the particle are different.
Of course, the explicit form of the coefficients
$\alpha_n^{(p)}$ are different for different models.
In what follows we restrict ourselves to the symmetric problem
when the transition rate is given by Eq.(\ref{W}).
In this case,
according to (\ref{alpha}) and (\ref{scaling+}), one  gets
\begin{eqnarray}
\alpha_n(V)=\frac{\nu}{4}\int d\xi \xi^n|\xi|\,f_M(V+\xi/2)
\nonumber
\end{eqnarray}
and therefore
\begin{eqnarray}
\alpha_n^{(p)}=\frac{\nu}{4}\int d\xi \xi^n|\xi|\,f_M^{(p)}(\xi/2).
\label{alpha_np}
\end{eqnarray}
Here the Maxwellian distribution for the bath molecules is
\begin{eqnarray}
f_M(v)=\left(\frac{\sigma}{2\pi}\right)^{1/2}
\exp\left(-\frac{1}{2}\sigma v^2\right),\,\,\,\,\,\,\,
\sigma=\frac{m}{k_BT},
\end{eqnarray}
and $f_M^{(p)}(\xi/2)=\frac{d^p}{dv^p}f_M(v)|_{v=\xi/2}$.

One can observe that $\alpha_n^{(p)}=0$
if $n+p$ is odd. Then the equation (\ref{myexpansion})
contains only operators of even indices $D_{2k}$,
which involve  $\alpha_n^{(p)}$ only with
even $n+p$. These coefficients  
$\alpha_n^{(p)}$ can be expressed
as linear combinations of the integrals
$
I_{k}=\int_{-\infty}^{\infty} d\xi\,\xi^{2k}\,|\xi|\,f_M(\xi/2)
$.
Using the first three of them 
\begin{eqnarray}
I_1=\frac{64}{\sqrt{2\pi}}\,\sigma^{-3/2},\,\,\,\,\,\,
I_2=\frac{1024}{\sqrt{2\pi}}\,\sigma^{-5/2},\,\,\,\,\,
I_3=\frac{24576}{\sqrt{2\pi}}\,\sigma^{-7/2}
\end{eqnarray}
one can calculate all coefficients $\alpha_n^{(p)}$
appearing in the equation of order $\lambda^6$, 
\begin{eqnarray}
\alpha_1^{(1)}&=&
-\frac{\nu\sigma}{8}I_1=-a\sigma^{-1/2},\nonumber\\
\alpha_1^{(3)}&=&\frac{3\nu\sigma^2}{8}I_1-
\frac{\nu\sigma^3}{32}I_2=
-a\,\sigma^{1/2},\nonumber\\
\alpha_1^{(5)}&=&-\frac{15\nu\sigma^3}{8}I_1
+\frac{5\nu\sigma^4}{16}I_2
-\frac{\nu\sigma^5}{128}I_3=
a\sigma^{3/2},\nonumber\\
\alpha_2^{(0)}&=&\frac{\nu}{4}I_1=
2a\sigma^{-3/2},\nonumber\\
\alpha_2^{(2)}&=&-\frac{\nu\sigma}{4}I_1+
\frac{\nu\sigma^2}{16}I_2=
6a\sigma^{-1/2},\nonumber\\
\alpha_2^{(4)}&=&\frac{3\nu\sigma^2}{4}I_1
-\frac{3\nu\sigma^3}{8}I_2
+\frac{\nu\sigma^4}{64}I_3=
6a\sigma^{1/2},\nonumber\\
\alpha_3^{(1)}&=&-\frac{\nu\sigma}{8}I_2=
-16a\sigma^{-3/2},\nonumber\\
\alpha_3^{(3)}&=&\frac{3\nu\sigma^2}{8}I_2
-\frac{\nu\sigma^3}{32}I_3=
-48a\sigma^{-1/2},\nonumber\\
\alpha_4^{(0)}&=&\frac{\nu}{4}I_2=
32a\sigma^{-5/2},\nonumber\\
\alpha_4^{(2)}&=&-\frac{\nu\sigma}{4}I_2
+\frac{\nu\sigma^2}{16}I_3=
160a\sigma^{-3/2},\nonumber\\
\alpha_5^{(1)}&=&-\frac{\nu\sigma}{8}I_3=
-384a\sigma^{-5/2},\nonumber\\
\alpha_6^{(0)}&=&\frac{\nu}{4}I_3=
768a\sigma^{-7/2}.
\end{eqnarray}
In these formulas  $a=\frac{8\nu}{\sqrt{2\pi}}$.

\section{Fokker-Planck and higher order equations}
The results of the previous section allow to write
the van  Kampen expansion 
for the Rayleigh model in the explicit form up to order $\lambda^6$.
Truncation of  (\ref{myexpansion}) to terms of order $\lambda^2$ leads to the
second order Fokker-Planck equation
\begin{eqnarray}
\frac{\partial f(x,t)}{\partial t}=\lambda^2D_2f(x,t),
\label{FPE}
\end{eqnarray}
where
\begin{eqnarray}
D_2=\frac{8\nu}{\sqrt{2\pi}}\Bigl\{
\sigma^{-1/2}\frac{\partial}{\partial x}x
+\sigma^{-3/2}\frac{\partial^2}{\partial x^2}\Bigr\},
\end{eqnarray}
and $\sigma=m/k_BT$.
It was shown recently that the  equation (\ref{FPE}),
first derived by Rayleigh, can be  recovered within a 
more general approach expressing coefficients in the 
$\lambda$-expansion in terms of 
correlation functions for the random force
and then taking the Markovian limit~\cite{PS}. 

The next approximation is the equation of order $\lambda^4$,
\begin{eqnarray}
\frac{\partial f(x,t)}{\partial t}=
\Bigl\{\lambda^2D_2+\lambda^4D_4\Bigr\}f(x,t),
\label{Eq4}
\end{eqnarray}
with
\begin{eqnarray}
D_4&=&\frac{8\nu}{\sqrt{2\pi}}\Bigl\{
-\sigma^{-1/2}\frac{\partial}{\partial x}x+
\frac{1}{6}\sigma^{1/2}\frac{\partial}{\partial x}x^3
-2\sigma^{-3/2}\frac{\partial^2}{\partial x^2}
+\frac{3}{2}\sigma^{-1/2}\frac{\partial^2}{\partial x^2}x^2
\nonumber\\
&&+\frac{8}{3}\sigma^{-3/2}\frac{\partial^3}{\partial x^3}x
+\frac{4}{3}\sigma^{-5/2}\frac{\partial^4}{\partial x^4}\Bigr\}.
\end{eqnarray}
This equation is equivalent to that obtained  
by van Kampen in~\cite{Kampen_paper}.

The equation of order $\lambda^6$ reads as
\begin{eqnarray}
\frac{\partial f(x,t)}{\partial t}=
\Bigl\{\lambda^2D_2+\lambda^4D_4+\lambda^6D_6\Bigr\}f(x,t),
\label{Eq6}
\end{eqnarray}
where
\begin{eqnarray}
D_6&=&\frac{8\nu}{\sqrt{2\pi}}\Bigl\{
\sigma^{-1/2}\frac{\partial}{\partial x}x
-\frac{1}{6}\sigma^{1/2}\frac{\partial}{\partial x}x^3
-\frac{1}{120}\sigma^{3/2}\frac{\partial}{\partial x}x^5
+3\sigma^{-3/2}\frac{\partial^2}{\partial x^2}\\
&&-3\sigma^{-1/2}\frac{\partial^2}{\partial x^2}x^2
+\frac{1}{8}\sigma^{1/2}\frac{\partial^2}{\partial x^2}x^4
-8\sigma^{-3/2}\frac{\partial^3}{\partial x^3}x
+\frac{4}{3}\sigma^{-1/2}\frac{\partial^3}{\partial x^3}x^3
\nonumber\\
&&-\frac{16}{3}\sigma^{-5/2}\frac{\partial^4}{\partial x^4}
+\frac{10}{3}\sigma^{-3/2}\frac{\partial^4}{\partial x^4}x^2
+\frac{16}{5}\sigma^{-5/2}\frac{\partial^5}{\partial x^5}x
+\frac{16}{15}\sigma^{-7/2}\frac{\partial^6}{\partial x^6}\nonumber
\Bigr\}.
\end{eqnarray}
This equation appeared previously in an incomplete form in~\cite{Colin}
missing the term 
$\alpha_5^{(1)}\frac{\partial^5}{\partial x^5}x$ in $D_6$. 

\end{widetext}

One can show~\cite{Kampen_paper2} that the stationary solution of the 
master equation with transition rates in the form (\ref{W})
is the Maxwellian distribution which for the scaled velocity
$x=\lambda^{-1} V$ has the form
$f_s(x)=C\exp\left(-\frac{1}{2}\sigma x^2\right)$.
This distribution does not depend on
$\lambda$ and therefore must
satisfy each term in the expansion (\ref{myexpansion}) 
separately,
\begin{eqnarray}
D_nf_s(x)=0.
\end{eqnarray}
One can immediately check that the above expressions for
$D_2$, $D_4$, and $D_6$ do satisfy this condition.

Eqs. (\ref{Eq4}) or (\ref{Eq6}) can be solved
using an appropriate perturbation technique, however
it is often easier to handle  
the equations for the moments 
$\langle x^n\rangle$ \cite{Kampen_book,Colin}.

It is known that approximations of the master equation by 
a differential equation involving derivatives of order higher
than two may lead to solutions which are not positive 
definite~\cite{Risken,Hanggi1}.
The reason why the Fokker-Planck equation (\ref{FPE})
preserves positivity of $f$ is 
because  the right hand side of the equation, 
and therefore the time derivative $f_t$,  are always positive at the points where
$f(x,t)$ as a function of $x$ has minima.
As a result, the minima become less deep with time,
the initially positive solution remains positive for all times. 
This is not true in general for 
Eqs. (\ref{Eq4}) and (\ref{Eq6})
involving $x$-derivatives
of order higher than two. These derivatives can be of any sign at extreme
points, which in principle may result in negative $f_t$
at minimum points.
However, since the terms with higher order derivatives 
are of higher order in $\lambda$,
one may expect that for sufficiently smooth initial distributions
the sign of the right sides of 
Eqs.~(\ref{Eq4}) and (\ref{Eq6})
is determined by the term with $f_{xx}$.
For example,
the only terms in the 
right side of Eq.(\ref{Eq4}) which can be negative at minimum points are
those involving $\lambda^4\sigma^{-3/2}xf_{xxx}$ and 
$\lambda^4\sigma^{-5/2}f_{xxxx}$.
They  are smaller than  the term 
$\lambda^2\sigma^{-3/2}f_{xx}$, which gives the positive
contribution, by the factors of order
$\lambda^2(x/x_c)$ and
$\lambda^2(x_{th}/x_c)^2$, respectively.
Here $x_c$ is the characteristic length
of the distribution $f(x,t)$ for a given $t$, and
$x_{th}$ is the scaled thermal velocity of the particle,
$x_{th}=\lambda^{-1}\sqrt{k_BT/M}$.
Recall that the expansion method we used implies
that the system is close to equilibrium and that
$x\sim x_{th}\sim\lambda^0$. 
Under this condition the term with $f_{xx}$ dominates
and determines the 
sign of the right hand side of Eq. (\ref{Eq4})  at minimum points, 
which guarantees 
preservation of the positivity of the solution.

The expansion method 
may be implemented also for
non-equilibrium fluctuations~\cite{non_equilibrium} 
but in that case the possibility
to introduce a stochastic variable $x$, which would be of order 
one for all relevant times, 
is less obvious and  has to be justified {\it a posteriori}.

\section{Discussion}
If the white random force $F(t)$ in the Langevin equation 
$\dot V(t)=A(V)+F(t)$ 
is also assumed to be a Gaussian process, then
the conventional procedure outlined in Section 2 
leads invariably to the second order Fokker-Planck equation~(\ref{FPE0}),
no matter whether the  damping force $A(V)$ is linear or not.
On the other hand, the van Kampen method, which does not require
any assumptions about
statistics of the random force,
leads to the second order equation
only to the lowest order in the expansion parameter $\lambda$, 
when the damping force is linear $A(V)= -\gamma V$.
In higher orders in $\lambda$, when the damping force $A(V)$ involves
nonlinear corrections $A(V)=-\gamma_1 V-\gamma_3 V^3-\cdots$ (where
$\gamma_3/\gamma_1\sim \lambda^2$), 
the VKE leads to equations with $V-$derivatives of order higher than two. 
This means  that {\it the approximation of a Gaussian random force
is legitimate to lowest order in $\lambda$ but not to higher orders.}
It is therefore inconsistent to take into account
nonlinear corrections to the linear  damping force $A(V)=-\gamma V$, 
and at the same time to assume that the random force
is Gaussian. Failure to appreciate this point may lead to
wrong conclusions. For example, 
for the Rayleigh model the Maxwell distribution is a correct
stationary solution to any order in $\lambda$. On the other hand
the assumption of a Gaussian random force 
would lead 
to the Fokker-Planck equation (\ref{FPE0}) which 
for nonlinear $A(V)$ has 
a non-Maxwellian stationary solution.

The Gaussian property of the random force 
is usually expected to hold interpreting the 
force as a result of many uncorrelated collisions, and appealing to the
the central limit theorem. However, one has to keep in  mind that
a decomposition of the total force exerted on the particle into 
a regular damping and ``random'' parts is a purely mathematical procedure,
which in general can be performed  with an appropriate projection operator
technique. The random force $F(t)$, obtained in this way
generally can not be interpreted as a superposition of many
``physical'' forces.
Moreover, it is not even a dynamical 
variable: its evolution in time is not governed  by
the Newtonian propagator $\exp(tL)$ with the Liouville operator $L$,
but by a  more complicated ``projected'' propagator. One can get
some insight in properties of the random force expanding it
in powers of $\lambda$ which has the form~\cite{PS}
\begin{eqnarray}
&&F(t)=F_0(t)+\lambda\int_0^t dt_1 S(t-t_1)F_0(t_1)\\
&&+\lambda^2\int_0^t dt_1\int_0^{t_1} dt_2
S(t-t_1)S(t_1-t_2)F_0(t_2)+\cdots\nonumber
\label{exx}
\end{eqnarray}
where $S(t)$ is a non-Newtonian propagator the explicit form of which is not
important for our purpose here. The first term $F_0(t)$ in the expansion
(\ref{exx}) is a dynamical
variable and has a well defined physical meaning: it is a force
exerted on the particle fixed in space (the limit of infinitely heavy 
particle). Conventional  qualitative reasoning to justify the  
Gaussian property is quite applicable for this term.  
Moreover, for a large Brownian particle, interacting
simultaneously with many bath molecules, the Gaussian property of
$F_0(t)$  can be proved analytically~\cite{PS}. To lowest order in $\lambda$
the random force is just $F_0(t)$ and therefore is Gaussian.
On the other hand, there is no  reason to expect that the same argument 
should work for the 
``unphysical forces'' represented in~(\ref{exx})
by terms  of higher orders in $\lambda$ involving the non-Newtonian 
propagator.  These terms in general are not Gaussian. 
In a future publication we shall  explicitly evaluate
correlation functions $\langle F(t_1)F(t_2)\cdots F(t_k)\rangle$
for an exactly solvable model with parabolic interaction
suggested in~\cite{PS}.
This would allow to construct the expansion similar to the VKE
but expressing coefficients in terms of
correlation functions for the random force rather than 
transition rates $W$. 
In lowest order in $\lambda$ it was done in~\cite{PS}
recovering in the Markovian limit
the Fokker-Planck equation (\ref{FPE}).

This work was supported by a grant from the NSERC.



\begin{thebibliography}{9}
\bibitem{B1} R. D. Astumian and P. H\"anggi, Phys. Today 55, 33
  (2002); J. Luczka, Physica A 274, 200 (1999);
P. Reimann and P. H\"anggi, Appl. Phys. A 75, 169 (2002).
\bibitem{Gruber} 
Ch. Gruber and J. Piasecki, Physica A 268, 412 (1999);
E. Kestemont, C. Van den Broeck, and M. M. Mansour, 
Europhys. Lett 49, 143 (2000);
T. Munakata and H. Ogawa, Phys. Rev. E 64, 036119 (2001).
\bibitem{I}
A.V. Plyukhin and J. Schofield, Phys. Rev. E 69, 021112 (2004).
\bibitem{Broeck2} P. Meurs, C. Van den Broeck, and A. Garcia,
Phys. Rev. E 70, 051109 (2004).
\bibitem{Res} L. Gammaitoni, P. H\"anggi, P. Jung, and F. Marchesoni,
Rev. Mod. Phys. 70, 223 (1998).
\bibitem{Risken} H. Risken, {\it The Fokker-Planck Equation},
(Springer, Berlin, 1989).
\bibitem{Kampen_book} N. G. van Kampen, {\it Stochastic Processes in Physics
and Chemistry}, (North Holland, Amsterdam, 1992).
\bibitem{Kampen_paper}
N. G. van Kampen, Can. J. Phys. 39, 551 (1961).
\bibitem{gen} J. L. Lebowitz and E. Rubin, Phys. Rev. 2381 (1964);
P. Resibois and H. T. Davis, Physica 30, 1077 (1964);
H. Mori, H. Fujisaka, and H. Shigematsu, Prog. Theor. Phys. 51, 109 (1974).
\bibitem{Barkai} E. Barkai,
J. Stat. Phys. 115, 1537 (2004).
\bibitem{PS} A.V. Plyukhin and J. Schofield, Phys. Rev. E 68, 041107
(2003).
\bibitem{non_equilibrium}
R. F. Fox and M. Kac, Biosystems 8, 187 (1977);
R. F. Rodriguez and L. S. Garcia-Colin, J. Phys. A 15, 527 (1982).
\bibitem{Hanggi2}
P. H\"anggi, H. Grabert, P. Talkner, and H. Thomas,
Phys. Rev. A 29, 371 (1984).
\bibitem{Colin}  R. F. Rodriguez and L. S. Garcia-Colin,  Phys. Lett. A
68, 151 (1978).
\bibitem{Zhu}S.-B. Zhu, Phys. Rev. A 42, 3374 (1990).
\bibitem{Hanggi1} P. H\"anggi and P. Talkner, J. Stat. Phys. 22, 65 (1980).
\bibitem{Kampen_paper2}
C.T.J.  Alkemade, N.G. van Kampen, and D.K.C. MacDonald, 
Proc. R. Soc. 271 A, 449 (1963).

\end{thebibliography}
\end{document}